\documentstyle[prl,aps,epsf]{revtex}
\draft
\begin{document}
\twocolumn[\hsize\textwidth\columnwidth\hsize\csname @twocolumnfalse\endcsname

\title{Noise at the Wigner Glass Transition} 
\author{C. Reichhardt and C.J. Olson Reichhardt} 
\address{ 
Center for Nonlinear Studies and Theoretical
Division, 
Los Alamos National Laboratory, Los Alamos, New Mexico 87545}

\date{\today}
\maketitle
\begin{abstract}
Using a simple model for interacting electrons in two dimensions
with random disorder, we show that a crossover from a Wigner
liquid to a Wigner glass occurs as
a function of charge density. 
The noise power increases strongly at the transition
and the characteristics of 
the $1/f^{\alpha}$ noise change.  
When the temperature is increased, the noise power decreases.
We compare these results
with recent noise measurements  
in systems with two-dimensional metal-insulator transitions. 
\end{abstract}
\vspace{-0.1in}
\vspace{-0.3in}

\vskip2pc]
\narrowtext
In the two-dimensional (2D) 
metal-insulator transition (MIT) regime, both the 
Coulomb interactions between electrons 
and the disorder are expected to be strong, 
leading to the formation of an electron glass \cite{Review,Rice}. 
Recent experiments in 2D electron systems have revealed
changes in the characteristics and the amplitude
of the conduction noise as the change density of the system
is varied from a high charge density metallic phase to a
lower charge density insulating phase
\cite{Popovic,Sn,Noise}. This
has been interpreted as evidence for an onset of glassy
dynamics near the insulating phase.  
These studies also find that the noise has 
a $1/f^{\alpha}$ characteristic, with $\alpha = 1.0$ in the metallic phase
changing over to 
$\alpha \approx 1.8$ in the glassy phase.
Other experiments have found similar results with $\alpha = 0.75$ in the
metallic phase and $\alpha =1.3$ near the insulating phase \cite{N2d}. 
The glassy phase is signified by a large increase in the
noise power along with the change in $\alpha$ \cite{Popovic,Sn}.
In addition, the noise power was observed to {\it decrease} 
with temperature \cite{Sn,Noise}, in contrast to single electron models
with thermally activated trapping \cite{Weissman} and
other models \cite{Voss} 
that predict an increase in the noise power with $T$. 
This suggests the importance of electron-electron
interactions at the MIT. 
Other recent experiments  
near the 2D MIT have also found
$1/f^{\alpha}$ noise, strong increases in noise power with decreasing 
charge density, and decreasing noise with increasing $T$ \cite{N2d,Tourbot}.  
Additionally, $1/f^{\alpha}$ noise fluctuations in thin granular films
have been interpreted as evidence for a glassy electron state
\cite{Wu}.

In theoretical studies, it was proposed  
that at the 2D MIT
a freezing from an electron liquid to  
a partially ordered Wigner glass \cite{Kivelson}
or a more strongly disordered electron glass \cite{Thakur,Glass} may occur. 
Other theories suggest that an intermediate metallic glass phase 
appears between the liquid and insulating phases \cite{Pastor}. 
It is also possible that the metallic 
glass phase may consist of
solid phase insulating regions coexisting
with string-like liquid regions.  
Studies of glassy systems 
often find cooperative string-like motions or dynamical 
heterogeneities \cite{Glotzer}. 
Such motions can give rise to correlated dynamics 
and large fluctuations near a glass transition. 
It is, however, unclear what the origin of such 
cooperativity would be in the electron glass systems. 

The strong enhancement of the noise observed
when passing
from a liquid to a glassy phase
in the 2D charge system has noticeable similarities to 
the noise found in studies of
vortex matter in superconductors \cite{Higgins}. 
In the vortex system, the existence of a transition
from a weakly pinned liquid phase to a more
strongly pinned glassy phase has been
established.
Noise studies in low temperature superconductors have shown that, 
as the glassy phase is approached from 
the liquid state, 
the voltage noise has a $1/f^{\alpha}$ characteristic with
increasing $\alpha$, corresponding to an increase 
in the noise power \cite{Higgins}. 
 
The vortex system studies
suggest that similar physics may be occurring in the 2D electron systems
when there is a transition from a liquid like phase with low noise power to 
a pinned glassy phase with high noise power. 
This is  
also consistent with the theoretical prediction that the electron liquid 
freezes
into a 2D disordered solid. 
In this work we propose a simple model for a classical 2D electron system
consisting of interacting electrons with random disorder and
temperature. We monitor the fluctuations and noise 
characteristics of the
current as a function of electron density or temperature. 
The advantage of our model is
that a large number of interacting electrons 
can be conveniently simulated, while a
full quantum mechanical model of similar size would be computationally 
prohibitive. 
Despite the limitations of this model, we show that this approach 
captures many of the key experimental observations.
Additionally, although our primary focus is to gain insight into the 
physics near the 2D MIT, our model is also relevant for other
classical charge systems undergoing 
transitions from glass to liquid states, such as charged
colloids interacting with random disorder.
Our model is similar to previous studies of 2D classical
electron systems with disorder \cite{Fertig,Reichhardt};
however, these previous studies focused 
on the microscopics of the defects in the lattice \cite{Fertig} 
or the sliding dynamics \cite{Reichhardt}. 
In the present work we focus on the 
noise fluctuations in the strongly disordered phase as it
changes from a liquid to a frozen state
as a function of electron density
for a fixed amount of disorder.

Our model consists of a 2D system of $N_{s}$ interacting electrons
with periodic boundary conditions in the $x$ and $y$-directions. 
There are also $N_{p}$ defect sites which 
attract the electrons. We assume the electron motion is at
finite temperature and the time evolution occurs through Langevin
dynamics. The damping on the electrons comes 

\begin{figure}
\center{
\epsfxsize=3.5in
\epsfbox{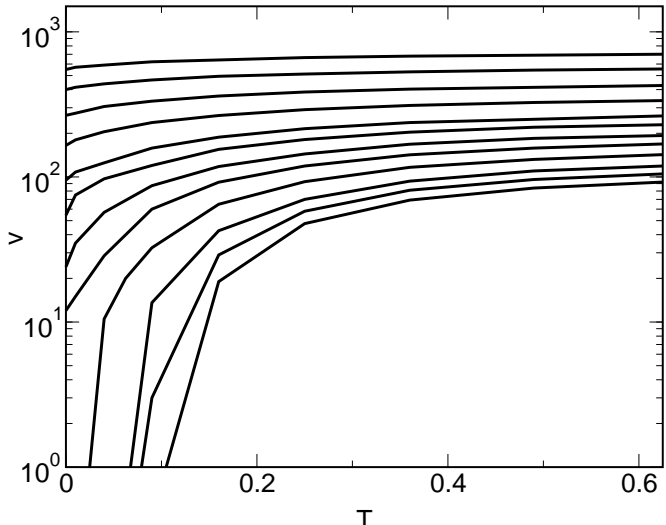}}
\caption{
The average electron velocity $v$ vs $T$ for 
$N_{s}/N_{p} = 2.67, 2.13, 1.67, 1.33, 1.07,
0.94, 0.8, 0.7, 0.6, 0.5, 0.45$, and $0.4$, from 
top to bottom. 
}
\end{figure}

\noindent
from their interactions
with phonons or small scattering sites.  
The equation of motion for a electron $i$ is 
\begin{equation}
\eta {\bf v}={\bf f}_{i}=-\sum_{j}^{N_{s}}\nabla U(r_{ij}) + 
{\bf f}_{i}^{s} + {\bf f}_{i}^{T} + {\bf f}_{d} 
\end{equation}
Here $\eta= 1$ is the damping constant and $U_{i}(r) = -q^2/r$,
with $q=1$, is the
electron-electron interaction potential, treated as in \cite{Jensen}.
The term ${\bf f}_{i}^{s}$ comes from the $N_p$ randomly spaced defect sites
modeled as 
parabolic traps of radius $r_{p}=0.2$ and strength $f_{p} = 1.0$. 
The
thermal noise ${\bf f}_{i}^{T}$ arises from random Langevin kicks
with 
$<f^{T}(t)> = 0$ and $<f^{T}_i(t)f^{T}_j(t^{\prime})> = 2\eta k_B T \delta_{ij}
\delta(t - t^{\prime})$. 
The driving term ${\bf f}_{d}=f_d \hat x$ comes from an applied voltage,
and we take $f_d= 0.1$.  
We start at a high temperature where the charges are diffusing rapidly and cool
to a lower temperature. 
We then wait for $10^4$ simulation time 
steps to reduce transient 
effects before appling the drive and measuring the average velocity 
$v$ of the electrons, which is
proportional to the conduction or resistance. 
We do this for a series of electron
densities at fixed disorder strength. We have considered samples
with constant pin densities $n_{p}$ for different system sizes 
such that $N_{p}$ ranges from $317$ to $1200$. 

We first consider the average 
electron velocity 
as a function of temperature and charge density
for a system with a fixed $N_{p}=1200$.
In Fig.~1 we show a series of conductance curves vs $T$ 
for charge density varied over nearly an 
order of magnitude, $0.4 \leq N_{s}/N_{p} \leq 2.67$. 
For high $N_{s}/N_{p}> 0.6$ the conductance is finite 
down to $T = 0$, while for $N_{s}/N_{p} \leq 0.6$,  
the electron velocity drops
to zero within our resolution, indicating that all the electrons 
are strongly pinned in an insulating phase.    
As $N_{s}/N_{p}$ increases above 1.0, the 
downward curvature of $v$ at low $T$ decreases.
These curves appear very similar 

\begin{figure}
\center{
\epsfxsize=3.5in
\epsfbox{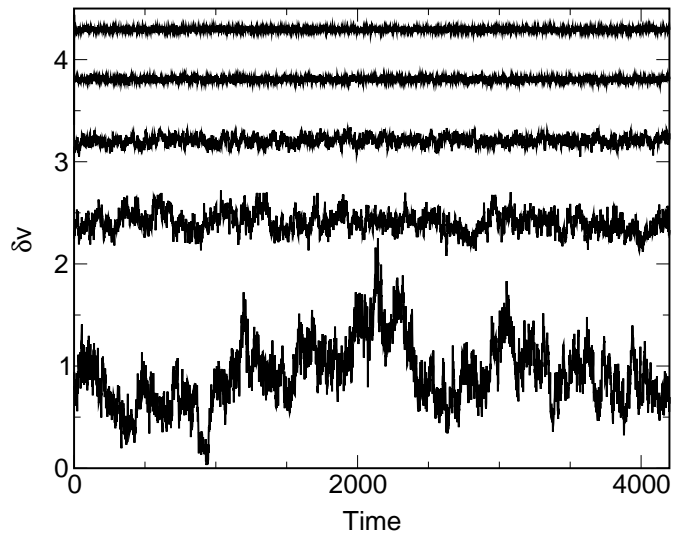}}
\caption{
The relative velocity fluctuations $\delta v$ vs time 
for the system in Fig.~1 for $T = 0.09$ at 
$N_{s}/N_{p} = 0.5, 0.7, 1.05, 1.64$, and
$2.67$, from bottom to top. The curves have been shifted up for clarity. 
}
\end{figure}

\noindent
to those typically observed in
2D MIT studies \cite{Popovic}. One difference is that
we do not find a charge density $n_{s}^{up}$ 
above which the slope of the velocities
turns up slightly at low $T$, as in the experiments. This may be due to 
the fact that in our model we do not directly include phonons. 
In the experimental regime of interest to us here, the large noise increases 
occur at charge densities $n_s<n^{up}_s$, where the 
velocity curves bend down at low temperature.

We next consider the relative fluctuations in the
velocities, $\delta v(t)=(v(t) - <v>)$ 
for varied $N_{s}/N_{p}$ at a fixed $T = 0.09$ for the system in Fig.~1.
This analysis is similar to that performed in experiments
\cite{Popovic,Sn,Noise}. 
In Fig.~2 we show the time traces of the relative 
velocity fluctuations for $N_{s}/N_{p} = 0.5, 0.7, 
1.05, 1.64$ and $2.67$. 
Here the fluctuations increase as $N_s$ drops,
in agreement with the experiments \cite{Popovic}. For 
$N_{s}/N_{p} < 0.44$,  the system is pinned and there are no fluctuations.
It is possible that, over a longer time interval such as that accessible
experimentally, there would be
even larger fluctuations at these small $N_s$ values; however, this is beyond
the time scale we can access with simulations.
A similar series of time traces can be obtained
for $\delta v$ at fixed $N_{s}/N_{p}$ for increasing temperature (not shown). 
Here the fluctuations are reduced at the higher temperatures,
in agreement with experiments \cite{Popovic}.

From the fluctuations $\delta v$ we measure the power spectrum  
\begin{equation}
S(\nu) = \left|\int \delta v(t)e^{-2\pi i\nu t}dt\right|^2  .
\end{equation}
In Fig.~3 we plot $S(\nu)$ for two different charge densities.
At $N_{s}/N_{p} = 0.5$ (upper curve), the spectrum shows a
$1/f^{\alpha}$ characteristic with $\alpha=1.37$ over a few
orders of

\begin{figure}
\center{
\epsfxsize=3.5in
\epsfbox{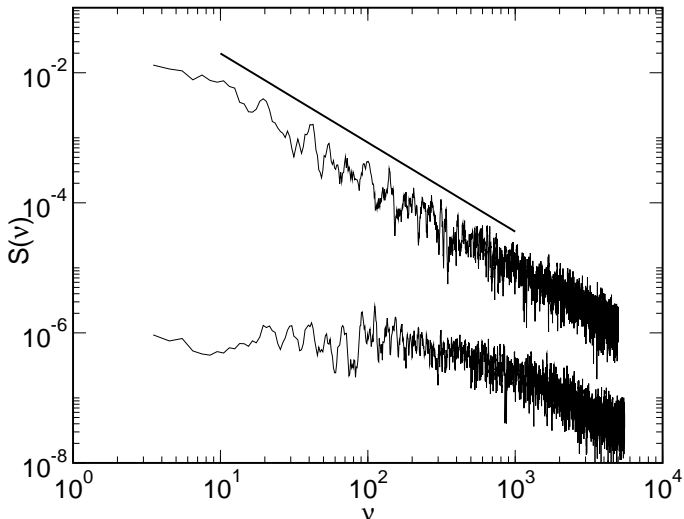}}
\caption{The power spectra $S(\nu)$ for the velocity fluctuations
$\delta v(t)$
at $N_{s}/N_{p} = 0.5$ (upper curve) and $N_{s}/N_{p} = 1.67$ (lower curve). 
The solid line has slope $\alpha = 1.37$.}
\end{figure}

\noindent
magnitude in the frequency. 
In contrast, for $N_{s}/N_{p} = 1.67$ (lower curve),
the noise power at lower frequencies is considerably reduced 
and the spectra is white
with $\alpha \approx 0$. We not that it is the lower frequcies which will be most
readly accesible in experiment. 
For fixed $N_{s}/N_{p} = 0.5$, we find that the power spectrum becomes
white upon increasing $T$.
We note that our results differ quantitatively from the experimental 
noise measurements \cite{Popovic} which find $1/f^{\alpha}$ noise
with $\alpha = 1.0$ in the metallic phase and $\alpha = 1.8$ in the  
glassy regime. 
Our exponent $\alpha=1.37$ is close to 
the $\alpha = 1.3$ found in the glassy regime in 
other experiments \cite{N2d}, where $\alpha=0.75$ in the metallic regime.
It is possible that the exponents are not universal
but depend on the details of the disorder strength; nevertheless, 
our results are
in qualitative agreement with the experiments. 

In Fig.~4(a) we show the noise power $S_{0}$  integrated over the 
first octave vs $N_{s}/N_{p}$ for a fixed 
$T = 0.09$.  The noise power increases by four orders of 
magnitude as $N_{s}/N_p$ is reduced. 
At low $N_{s}/N_{p}$, the noise power decreases almost
exponentially with charge density and begins to saturate
at high $N_{s}/N_p$. 
Both these observations are in agreement with
the experimental results \cite{Sn,Noise}.
In the inset of Fig.~4(a) we plot the noise spectrum exponent
$\alpha$ vs $N_{s}/N_{p}$.
A large increase in $\alpha$ occurs near $N_{s}/N_{p} = 0.7$.
A similar sharp increase in the 
exponent is also observed in experiments
\cite{Sn,Noise,N2d} as a function of charge density and has been 
interpreted
as the glassy freezing transition.
In Fig.~4(b) we plot $S_{0}$ vs $T$ for fixed $N_{s}/N_{p} = 0.5$. 
Here the noise power drops exponentially over four orders
of magnitude with increasing $T$, 
which is in agreement with the experiments \cite{Sn,Noise}.
We note that most single electron models predict
an {\it increase} in the noise power with temperature \cite{Weissman,Voss}.
Other models in the hopping regime predict a noise power that
is independent of $T$ or has a power law decrease with $T$ \cite{Kozub}. 
These discrepancies 

\begin{figure}
\center{
\epsfxsize=3.5in
\epsfbox{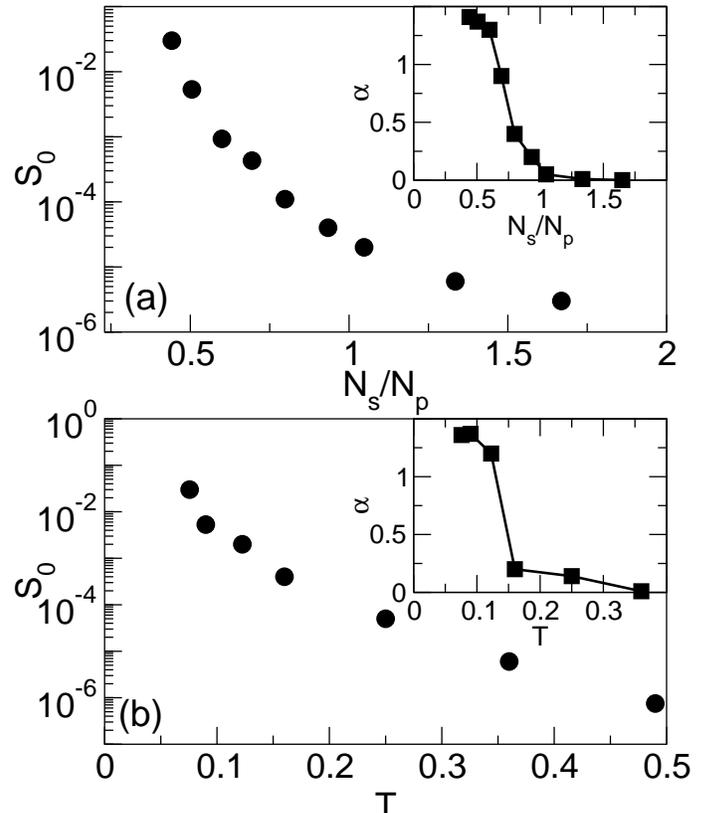}}
\caption{(a) The integrated noise power $S_{0}$ vs $N_{s}/N_{p}$
for $T = 0.09$. Inset: The power spectrum exponent $\alpha$ vs 
$N_{s}/N_{p}$ for $T = 0.09$. (b) $S_{0}$ vs $T$ for 
$N_{s}/N_{p} = 0.5$. Inset: $\alpha$ vs $T$ for
fixed $N_{s}/N_{p}=0.5$. 
}
\end{figure}

\noindent
suggest that the noise in the experiment is not
due to single electron hopping events, but is instead
caused by correlated electron motions. 
In the inset of Fig.~4(b) we plot $\alpha$ vs $T$, showing that a sharp
increase in $\alpha$ occurs near $T = 0.125$ at the onset of the
glassy freezing.

We have also measured the non-Gaussian nature of the
noise. At high $N_{s}$ and $T$ the noise fluctuations are
Gaussian; however, in the regions of high noise power we find 
non-Gaussian noise fluctuations with a skewed distribution. 
Experiments have also found evidence for non-Gaussian fluctuations
in the glassy regimes \cite{Noise}.

Next we show evidence that the large noise is due to correlated regions
of string like electron flow, and that within 
these regions the electrons move in 1D or quasi 1D channels. 
Because of the reduced dimensionality
the electron motion is more correlated. 
In Fig.~5(a) we show the trajectories of the electrons for a fixed period
of time for a system with $T = 0.09$ at $N_{s}/N_{p} = 1.67$. Here
the electrons can flow freely throughout the sample, although there are some
areas where electrons become temporarily trapped by a defect site. 
In Fig.~5(b) at $N_{s}/N_{p} = 1.37$,
where the noise power is larger than in the system shown in Fig.~5(a), 
larger pinned regions appear and the electron motion consists of a mixture of 
2D and 1D regions. If the trajectories are followed over
longer times, 

\begin{figure}
\center{
\epsfxsize=3.45in
\epsfbox{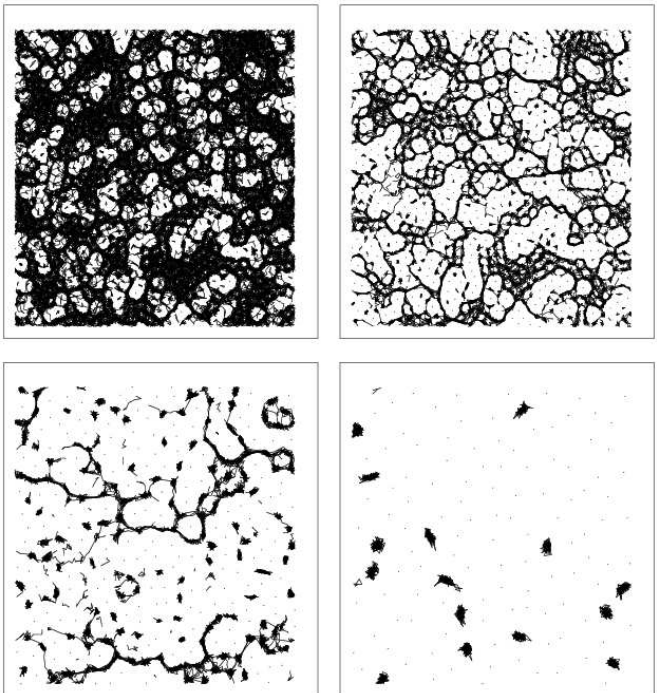}}
\caption{Electron trajectories for a fixed period of time for fixed 
$T = 0.09$ at (a)$N_{s}/N_{p} = 1.67$, (b) $1.37$, (c)$0.5$, 
and (d) $0.3$.}
\end{figure}

\noindent
motion occurs throughout the entire sample. 
In Fig.~5(c) at $N_{s}/N_{p} = 0.5$, where the noise power
and $\alpha$ are both maximum, 
the electron motion occurs mostly in the form
of 1D channels that percolate through the sample.
There are also regions where the electron motion
occurs in small rings.
The channel structures change very slowly with time, with
a channel occasionally shutting off while another emerges
elsewhere.  It is the intermittent opening of the 1D channels 
which gives rise to the large noise fluctuations in this regime. 
When a percolating 1D channel opens,
all the electrons in that channel move in 
a correlated fashion leading to a large increase in the conduction. 
Conversely,
if a percolating channel closes all the electrons
in that channel cease to move. It is well known that 
fluctuations in 1D are much more strongly enhanced than in 2D. As
$T$ or $N_{s}$ is increased, 
the motion 
becomes increasingly 2D in nature and the strong correlations 
of the electron motion are lost. 
We also note that the appearance of string like motions in the large
noise regions is consistent with studies in glassy systems,
where dynamical heterogeneities in the 
form of 1D stringlike motions of particles have been observed 
in conjunction with large noise \cite{Glotzer}. 
In Fig.~5(d) at $N_{s}/N_{p} = 0.3$, deep in the insulating regime, 
there are no channels. Instead, the 
infrequent motion of electrons occurs only by small jumps from defect
to defect. 

In conclusion, we have presented a simple model for
the glassy freezing of interacting electrons
in 2D with random disorder. 
For high electron density or high temperatures, the electrons form a 
2D liquid state and we find
low conduction noise power with a white spectra.
As the density of the electrons is lowered
for fixed temperature, or conversely, as
the temperature is lowered for fixed low electron density, there is a  
crossover to a $1/f^{\alpha}$ noise 
with large low frequency power and $\alpha = 1.37$.  
In this glassy regime, the electrons move in 1D intermittent 
stringlike paths which percolate throughout the sample.
Similar stringlike motions are also observed in other glass forming systems. 
For low electron density, all the electrons are frozen by the defect sites
and the motion occurs only by single electron hopping events. 
We find that the noise
power decreases exponentially with temperature, in agreement with
experiment. Many of our results are in qualitative agreement with recent 
experiments on 2D electron systems near the metal insulator transition. 

This work was supported by the US DoE under Contract No. W-7405-ENG-36.

\end{document}